\documentclass[twocolumn,aps,showpacs]{revtex4}
\usepackage{amsmath,psfrag,graphicx}

\def\vct#1{\mbox{\boldmath $#1$}}
\def\ket#1{|#1\rangle}

\begin{document}

\preprint{pra8.0}
\title{Absorption-free optical control of spin systems:\\
the quantum Zeno effect in optical pumping}
\author{T. Nakanishi, K. Yamane, and M. Kitano}
\affiliation{Department of Electronic Science and Engineering,
Kyoto University\\
Kyoto 606-8501, Japan}

\date{\today}

\begin{abstract}
\vspace{0.5cm}
We show that atomic spin motion can be controlled
by circularly polarized light
without light absorption in the strong pumping limit.
In this limit, the pumping light, which drives the
empty spin state, destroys the Zeeman coherence effectively
and freezes the coherent transition via the quantum Zeno effect.
It is verified experimentally that
the amount of light absorption decreases asymptotically to zero
as the incident light intensity is increased.
\end{abstract}

\maketitle

The quantum Zeno effect is known as the suppression 
of unitary time evolution caused by decoherence
in quantum systems \cite{Misra,Itano}.
The decoherence is provided by measurement,
interactions with the environment,
stochastic fields, and so on.
It is interesting in the sense that
the coherent, unitary evolution can be canceled
by decoherent processes in a probability conserving
manner.
If we look at the effect from the frame which moves
so as to cancel the unitary evolution,
the system otherwise stationary
is guided by the decoherence processes.
Using this inverse Zeno effect, we can control
quantum systems with incoherent processes
without the loss of probability\cite{Altenmuler}.

In this paper, we will show that the quantum Zeno effect plays
a crucial role in the optical pumping scheme.
Though not widely recognized, 
the light-induced decoherence has a strong influence on the spin dynamics.
Optical pumping is a useful method to polarize atoms \cite{Happer}.
Let us consider a four-level atom
with a $J=1/2$ ground state and a $J^\prime=1/2$ excited state.
In order to polarize the atomic spin in the ground state,
we apply a $\sigma_+$ pumping light beam which is propagated 
in the $z$ direction and  tuned to the transition 
between the ground state and the excited state.
The only spin-down atoms ($m_J=-1/2$) are excited by absorbing 
the pumping light,
and relax either to the spin-up state ($m_J=1/2$) or to the spin-down 
state in the ground level.
On the other hand the spin-up atoms cannot absorb the pumping light.
When the pumping light is strong enough, 
all the spin-down atoms are pumped out to the spin-up state.
Once the spins are completely polarized,
the atomic ensemble does not absorb the pumping light any more.

However in the presence of a magnetic field transverse to the pumping light,
the polarized spins are flipped back to the spin-down state.
The magnetic field tends to
equalize the population of the spin-down state with
that of the spin-up state through the precession.
Therefore the absorption of the pumping light is needed
to compensate the transition caused by the magnetic field.

But it has been found that the spin polarization 
can be retained against the transverse magnetic field
without any absorption of the pumping light \cite{Kitano}.
At first sight, it may seem contradictory,
but the phenomenon can be understood by considering the decoherence 
caused by the pumping light.
The spin precession itself is suppressed by the light-induced decoherence 
via the quantum Zeno effect,
which accompanies no light absorption.

In terms of the normal interpretation of the quantum Zeno effect
that the time evolution is suppressed by the
frequent measurements,
the transmission monitoring of the polarized light 
plays the role of measurements,
from which we can infer the spin direction.
The absence of absorption is clear evidence that all the spin are in the
up state.
By performing the null measurements frequently or continuously,
one can freeze the spin dynamics.
This is a kind of interaction-free measurement \cite{Kwiat}.

When we apply a $\sigma_+$ pumping light beam 
in the $z$ direction under an external 
magnetic field $\vct{B}_0$,
the normalized magnetic moment $\vct{m}$
obeys the Bloch equation:
\begin{align}
 \frac{ d \vct{m} } { d t } = \vct{m} \times \vct{\Omega}_0  
 - \Gamma \vct{m}  - P ( \vct{m}  - \vct{e}_z ),
 \label{bloch}
\end{align}
where $(\vct{e}_x , \vct{e}_y , \vct{e}_z )$ are the Cartesian unit vectors.
The first term of the right-hand side represents the spin precession 
around  $\vct{B}_0$ at the angular frequency 
$|\vct{\Omega}_0| = \gamma_g |\vct{B}_0|$
with the gyromagnetic ratio $\gamma_g$.
The second term represents the spin relaxation
by which the magnitude of the magnetic moment is decreased 
exponentially at the rate $\Gamma$.
The relaxation is not an essential element for the quantum Zeno effect
but we cannot ignore it in the actual experiment.
The third term represents the optical pumping toward
$\vct{m}=\vct{e}_z$ with the pumping rate $P$,
which is proportional to the light intensity $I$.

The steady-state solution to Eq.~(\ref{bloch}) is
\begin{align}
 \vct{m} = - \frac{ P \Omega_0 }{ ( P + \Gamma )^2 + \Omega_0^2 } \,
 \vct{e}_x + \frac{ P ( P + \Gamma )}
 { ( P + \Gamma )^2 + \Omega_0^2 } \, 
 \vct{e}_z , \label{steady}
\end{align}
where we have assumed that the magnetic field is applied
in the $y$ direction;
$\vct{\Omega}_0=\Omega_0\vct{e}_y$.

For the strong pumping limit,
i.e. for $P \gg \Omega_0$ , $P \gg \Gamma$,
we have $\vct{m} \sim  \vct{e}_z-O(\Omega_0/P) \vct{e}_x$.
This means that in this limit
all the spin are polarized in the $z$ direction
even in the presence of the transverse magnetic field.

The key point of the optical pumping is the selective
population transfer via excited states
and a population distribution
far from equilibrium in the ground states can be achieved.
Therefore one might think that in the above situation
the spin polarization is maintained by the repumping
(population transfer)
of the atoms flipped by the transverse magnetic field.
However, as will be shown below, this is not the case.
The spin flip itself is suppressed by the quantum Zeno
effect induced by the pumping light.
By exciting the empty state (the spin-down state in this case),
the pumping light destroys the Zeeman coherence and
results in the Zeno effect.

One can distinguish the two scenarios by observing the absorption
of the pumping light.
In the former case, 
we should see some absorption corresponding to
the repumping of the spin flipped by the magnetic field.
In the latter case, however,
we will see no absorption, at least in principle,
because the spin rotation is suppressed owing to the Zeno effect.
The Zeno effect can be realized without absorption when the
decoherence is provided by the excitation of the empty state.

We will separate the pumping term in the Bloch equation
(\ref{bloch}) into 
the pumping (population transfer) effect and the decoherence effect.
The pumping effect appears in the $z$ component as
\begin{align}
  \frac{d m_z}{d t} \Big|_{\rm pump} = P ( 1 - m_z ),
  \label{population} 
\end{align}
which represents the population transfer
from the spin-down state to the spin-up state.
The population transfer necessarily accompanies optical absorption
(and subsequent reemission).
For the completely polarized state,
$m_z \sim  1$,
this term has little effect on the spin dynamics.

On the other hand, the decoherence effect appears in
the $x , y$ components as
\begin{align}
  \frac{ d m_{x,y} }{d t} \Big|_{\rm pump} = - P m_{x,y}.
  \label{coherence}
\end{align}
Note that both $m_x$ and $m_y$ correspond to
the coherence $\rho_{+-}$ between 
the spin-up $\ket{+}$ and spin-down $\ket{-}$ states.
The coherence decays at the rate $\Gamma+P$ 
in the presence of the pumping light.
Therefore, Eq.~(\ref{coherence}) represents the decoherence.
This decoherence causes the suppression of the spin precession
via the quantum Zeno effect.
It does not accompany optical absorption and
works even when $m_z\sim 1$. 

The pumping rate $P$ is connected with
the pumping light intensity $I$ as
\begin{align}
 P = \eta \, \sigma \, \frac{I}{\hbar \omega}, \label{intensity}
\end{align}
where $\eta$, $\sigma$, and $\hbar \omega$ are the pumping efficiency,
the absorption cross section, and the photon energy respectively.
The absorption coefficient $\alpha$ is defined as
\begin{align}
 \alpha = N \, \sigma \, \frac{1}{2} ( 1 - m_z ),\label{alpha}
\end{align}
where $N$ is the atomic density.
Substituting $m_z$ in Eq.~(\ref{alpha}) by Eq.~(\ref{steady}),
we obtain
\begin{align}
 \alpha 
 = \frac{N \sigma}{2} \, \frac{ \Gamma ( \Gamma + P ) + \Omega_0^2 } 
 { ( \Gamma + P )^2 + \Omega_0^2 }. \label{rate}
\end{align}
For optically thin cases,
i.e. $\alpha \Delta z \ll 1$,
the absorption of the pumping light $\Delta I$ is given by 
$
 \Delta I =
 I \, \alpha \Delta z, 
$
where $\Delta z$ is the length of the cell.
With Eqs.~(\ref{intensity}) and (\ref{rate}),
we obtain
\begin{align}
  \Delta I = \epsilon P \,
  \frac{ \Gamma ( \Gamma + P ) + \Omega_0^2 } 
  { ( \Gamma + P )^2 + \Omega_0^2 } \, \Delta z, \label{absorption}
\end{align}
where we introduced $\epsilon=N \hbar \omega/2 \eta$ for simplicity.

\begin{figure}[t]
  \psfrag{pumping rate}[][bc]
 {{\bf{pumping rate $P/\Gamma$}}}
 \psfrag{absorption}[][tc]
 {{\bf{absorption $\Delta I/\epsilon \Gamma \Delta z$}}}
 \includegraphics[scale=0.6]{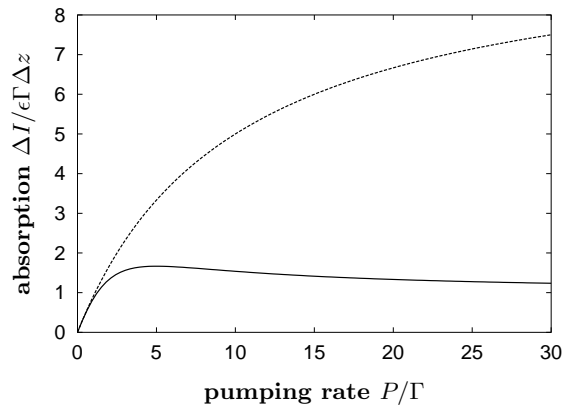}
 \caption{The solid line represents the theoretical absorption.
The dashed line is the absorption curve obtained by neglecting
the terms related to the quantum Zeno effect.
The optical pumping rate $P$ and the angular frequency of the precession
$\Omega_0$ is normalized by the relaxation rate $\Gamma$ ,
and these lines are calculated for $\Omega_0/\Gamma=3$.
The absorption rate is also normalized to satisfy 
$\lim_{P/\Omega_0 \rightarrow \infty} \Delta I/I_0 = 1$ .
}
 \label{graph}
\end{figure}

Here we see that $\Delta I$ monotonically increases 
with $P$ when $\Gamma > \Omega_0$, i.e.,
when the relaxation dominates the spin precession.
However,
for $\Gamma < \Omega_0$,
$\Delta I$ increases up to a point 
$ P = (\Omega_0^2 + \Gamma^2) / (\Omega_0 - \Gamma)$,
then $\Delta I$ starts to decrease.
In Fig.~\ref{graph} we plotted $\Delta I$ as a function of $P$
in the case of $\Gamma < \Omega_0$.
The absorption approaches asymptotically a specific value:
\begin{align}
 \lim_{P/\Omega_0 \rightarrow \infty} \Delta I 
 = \epsilon \Gamma \Delta z. \label{strong_pumping}
\end{align}

Thus, in the ideal condition where the intrinsic relaxation can be neglected,
one can suppress the spin evolution by the pumping light 
with no absorption.
Notice that Eq.~(\ref{strong_pumping}) is independent 
of the precession frequency $\Omega_0$
and is proportional to the relaxation rate $\Gamma$.
The relaxation is an incoherent process and
cannot be suppressed by the Zeno effect.
On the other hand, the precession is a coherent process and can be
suppressed.

The dashed line in Fig.~\ref{graph} represents the
imaginary absorption curve 
for which the decoherence presented by Eq.~(\ref{coherence})
is neglected factitiously.
The actual absorption (the solid line) is much smaller than
the dashed line.
Moreover,
the solid line is qualitatively different from the dashed line
in that it has a maximum point and decreases with $P$ thereafter.
If the spin precession were not suppressed by the pumping light,
we would have a normal saturated absorption profile 
like as the dashed line in Fig.~\ref{graph}.
This is because the polarized spins continue to precess at a certain rate
and absorb the pumping light.
The fact that the absorption decreases with $P$ is
a clear evidence for the suppression of the precession
by the pumping-induced decoherence. 
The reduction of the absorption is a peculiar feature
originating from the quantum Zeno effect.

\begin{figure}[b]
 \includegraphics[scale=0.5]{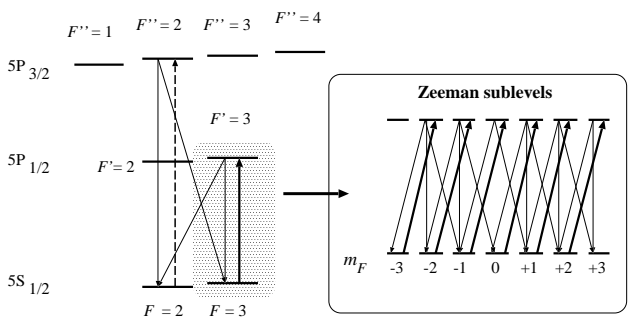}
 \caption{$\rm{^{85}Rb}$ energy diagram.
The arrows represent
optical pumping (thick lines),
repumping (dashed line),
and the spontaneous emission (solid lines).}
 \label{Rb85_transition}
\end{figure}

We use $\rm{^{85} Rb}$ atoms to verify the absorption-free
feature experimentally.
The energy diagram for $\rm{^{85} Rb}$ is illustrated 
in Fig.~\ref{Rb85_transition}.
A laser for the optical pumping is tuned to the transition frequency
between $F=3$ and $F^\prime=3$ in the $\rm{D_1}$ line ($\rm{5S_{1/2}}
\rightarrow \rm{5P_{1/2}}$).
The circularly polarized pumping light ($\sigma_+$) induces 
the transitions satisfying $\Delta m_F = +1$,
and eventually populates all the atoms to the Zeeman sublevel $m_F = 3$ 
in the absence of the transverse magnetic field.

The pumping cycle is not closed within the transition
between $F=3$ and $F^{\prime}=3$;
some of excited atoms in $F^{\prime}=3$ drop to the $F=2$ level
and are accumulated in that level.
In order to prevent the hyperfine pumping
we prepared another laser,
which was tuned to the transition frequency from the $F=2$ level
in the ground state to the $\rm{5P_{3/2}}$ state ($\rm{D_2}$ line). 
When the laser intensity is strong enough to repump the atoms
in the $F=2$ level,
the pumping cycle is effectively closed (meshed region).
The intensity of the repumping light must be comparable to or
,more than that of the pumping laser.

In our experimental condition,
the absorption coefficient for the pumping light 
is so small that we had to measure it with some precision.
We used a differential absorption method.
When the repumping light is turned on,
Rb atoms show the expected absorption of the pumping light.
On the other hand when the repumping light is turned off,
due to the hyperfine pumping,
there remains few atoms in the $F=3$ level,
and the absorption of the pumping light becomes negligibly small.
The difference in the transmitted intensity of the pumping light 
between the two cases
corresponds to the absorption $\Delta I$ to be measured.

\begin{figure}
 \includegraphics[scale=0.5]{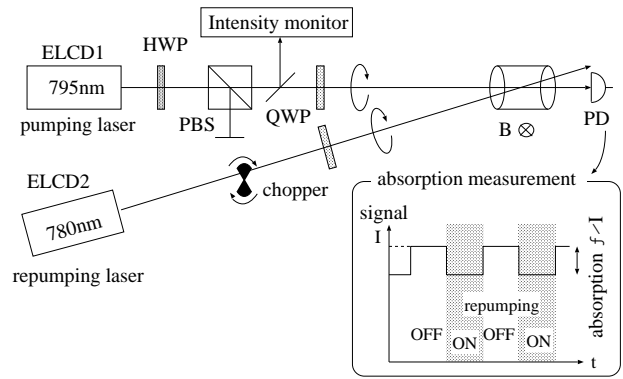}
 \caption{Experimental setup.
The inset shows the output signal of the photo detector (PD).
Chopping of the repumping laser results in
the rectangular wave whose amplitude corresponds
to the absorption of the pumping light.}
 \label{setup}
\end{figure}

Our experimental setup is illustrated in Fig.~\ref{setup}.
We used an external-cavity diode laser
with maximum output power 40\,mW for pumping ($\rm{D_1}$ line).
A similar system was prepared for repumping ($\rm{D_2}$ line).
The power of the pumping laser can be adjusted
using a half-wave plate and a polarizing beam splitter.
The pumping and repumping beams were aligned
so as to well overlap each other in the Rb cell.
The cell 
(2.2\,cm in length and 2.0\,cm in diameter)
is kept at room temperature.
The cross section of the pumping 
beam , $A_{\rm p}$, is about 0.12\,$\rm{cm^2}$,
and that of the repumping beam, $A_{\rm r}$, is about 0.36\,$\rm{cm^2}$.

It is important to minimize the spin relaxation.
The primary factor of the relaxation is
the diffusion of the atoms to the cell wall.
In order to suppress the atomic diffusion
we used nitrogen as buffer gas at 8.0\,kPa.
A uniform magnetic field can be applied to the Rb cell
by three pairs of Helmholtz coils.

As shown in Fig.~\ref{setup},
we measure the transmitted light intensity using a photo diode (PD)
while chopping the repumping laser by a mechanical chopper.
We have a rectangular waveform whose amplitude corresponds to the
absorption of the pumping light (see the inset).
The chopping frequency (91\,Hz) is smaller than 
the characteristic frequencies of the
system, $P$, $\Omega_0$ , and $\Gamma$ ( $>$ several kHz).

In order to increase the pumping rate effectively,
the transmitted light through the Rb cell is reflected back by a mirror
so that the reflected light takes almost the same path.
It should be noted 
that it doesn't mean only that the light path $\Delta z$ interacting with 
Rb atoms is doubled.
By overlapping the beams we can also doubles the pumping rate $P$ effectively.
Then Eq.~(\ref{intensity}) is modified as
\begin{align}
 P = \frac{2 \eta \sigma}{\hbar \omega} I, \label{intensity2}
\end{align}
and the absorption (\ref{absorption}) is replaced by
\begin{align}
 \Delta I &=  I \, \alpha \cdot 2\Delta z 
 = \epsilon P \,
 \frac{ \Gamma ( \Gamma + P ) + \Omega_0^2 } 
 { ( \Gamma + P )^2 + \Omega_0^2 } \, \Delta z. \label{absorption2}
\end{align}

\begin{figure}
 \psfrag{intensity}[][bc]
 {{\bf{pumping power $A_{\rm p} I / \rm{mW}$}}}
 \psfrag{absorption}[][tc]
 {{\bf{absorption $A_{\rm p} \Delta I / \rm{\mu W}$ }}}
 \includegraphics[scale=0.6]{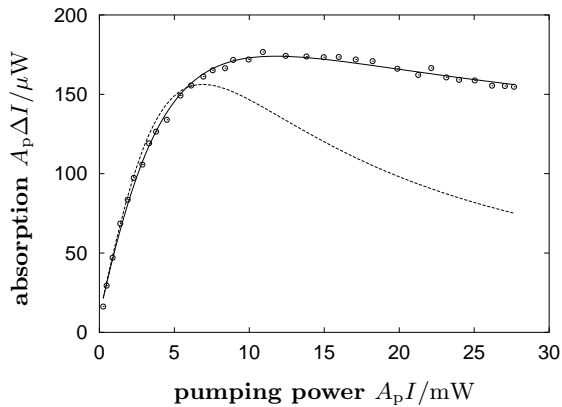}
 \caption{Experimental results ($\odot$),
theoretical line fitted with the results (solid line),
and theoretical line without relaxation (dashed line).
Applied magnetic field was 2\,$\mu$T.}
 \label{exp}
\end{figure}

Figure \ref{exp} shows the experimentally obtained absorption curve.
The power of the pumping laser incident to the Rb cell 
was varied from 0 to 28\,mW,
and the power of the repumping laser was fixed at 32.8\,mW.
The strength of applied magnetic field was 2\,$\mu$T
($\Omega_0 \sim 9\,\rm{kHz}$).
Fitting the results to the theoretical curve (solid line),
we estimated the relaxation rate to be about 3\,kHz.
Owing to the differential absorption technique,
we were able to measure small variation in the absorption constant
($\Delta I/I < 2\%$)
with high accuracy ( $<$ 0.1\%).

The result shows that the absorption decreases 
when the pumping power is over about 10\,mW.
The decrease in the absorption was more than fluctuations (a few $\mu$W).
We emphasize repeatedly that
the decrease in the absorption is qualitatively different 
either from linear absorption or from saturated absorption.
This phenomenon is induced by the Zeno suppression 
of the spin precession.
The theoretical curve shows good agreement with
the experimental result.

As mentioned before,
in the strong pumping limit
the spin precession can be suppressed completely 
by the quantum Zeno effect,
but the spin relaxation still persists.
The dashed line in Fig.~\ref{exp} shows the absorption curve 
for which the absorption related to the relaxation is subtracted.
This line declines more rapidly with the increase of pumping power,
and is supposed to tend to zero.
If we were to eliminate the intrinsic relaxation,
we would have the pronounced decrease in the absorption.

We can increase buffer gas pressure to reduce the relaxation 
due to the diffusion.
However the spin relaxation by the collision increases 
in stead,
and we can not expect drastic improvement.
A considerable improvement will be expected by cooling the atoms.
But the laser cooling is usually 
performed under inhomogeneous magnetic fields,
so we have to turn off the magnetic fields
as well as the cooling laser beams.
Another possible method is the use of a
glass cell 
containing dense $\rm{^4He}$ buffer gas.
At the temperature blow 2.1 K, the inner wall is
coated with liquid $\rm{^4He}$ films and
the spin relaxation time is significantly reduced \cite{Hatakeyama}.
This method has the advantage that it is free from
magnetic fields and gravity.
It is reported that the spin polarization time can be extended to
as long as $\sim 60 \,$s.

The suppression of the coherent transition by the optical pumping 
is applicable to the quantum-state control of highly
degenerate multilevel systems such as in quantum logic gates.
In order to prevent the population diffusion to undesired
levels, we can optically pump these empty levels selectively.
Moreover, we can attain active
control of spin states using the inverse Zeno effect.
By changing the
pumping direction or the polarization
continuously, we can change the spin direction arbitrarily.
This is similar to the coherent population trapping or the adiabatic
passage with the dark state \cite{Kitano2}. 
The recent paper \cite{Luis} presents the application to 
the preparation of quantum state in superposition.

We wish to thank T. Ikushima for helpful discussions.
This research was supported by the Ministry of
Education, Culture, Sports, Science and Technology
in Japan under a Grant-in-Aid for Scientific Research
No.~11216203 and No.~11650043.

\end{document}